\documentclass[journal]{IEEEtran}

\usepackage{cite,graphicx,amsmath,amssymb}
\usepackage{subfigure}
\usepackage{citesort}
\usepackage{fancyhdr}
\usepackage{mdwmath}
\usepackage{mdwtab}
\usepackage{balance}
\usepackage{xcolor}
\usepackage{bm}
\usepackage{cases}
\usepackage{algorithm}
\usepackage{algorithmic}
\usepackage{multirow}
\usepackage{flafter}
\usepackage{epstopdf}
\usepackage[justification=centering]{caption}

\newtheorem{remark}{\textbf{{\emph{Remark}}}}

\newtheorem{theorem}{\textbf{\emph{Theorem}}}

\begin{document}
	
	\title{Reconfigurable Intelligent Surface Enhanced NOMA Assisted Backscatter Communication System}
 	\author{ Jiakuo~Zuo, Yuanwei~Liu,~\IEEEmembership{Senior Member,~IEEE}, Liang Yang, Lingyang Song,~\IEEEmembership{Fellow,~IEEE}, and Ying-Chang Liang,~\IEEEmembership{Fellow,~IEEE} 
  \thanks{J. Zuo is is with the School of Internet of Things,  Nanjing University of Posts and Telecommunications, Nanjing 210003, China (e-mail: zuojiakuo@njupt.edu.cn).}
  	\thanks{Y. Liu is with the School of Electronic Engineering and Computer Science, the Queen Mary University of London, London E1 4NS, U.K. (e-mail: yuanwei.liu@qmul.ac.uk).}
  	 \thanks{L. Yang is with the College of Computer Science and Electronic Engineer-ing, Hunan University, 410082 China (e-mail: liangy@hnu.edu.cn).}
  	 \thanks{L. Song is with Department of Electronics, Peking University, Beijing, 100080 China (e-mail: lingyang.song@pku.edu.cn).}
  \thanks{Y.-C. Liang is with the Center for Intelligent Networking and Communications (CINC), University of Electronic Science
 	and Technology of China (UESTC), Chengdu 611731, China (e-mails:
 	liangyc@ieee.org).} 
}
	\maketitle
	\vspace{-1cm}
	\begin{abstract}
		A reconfigurable intelligent surface (RIS) enhanced  non-orthogonal multiple access assisted backscatter communication (RIS-NOMABC) system is considered. A joint optimization problem over power reflection coefficients and phase shifts is formulated. To solve this non-convex problem, a low complexity algorithm is proposed by invoking the alternative optimization, successive convex approximation and manifold optimization algorithms. Numerical results corroborate that the proposed
		RIS-NOMABC system outperforms the conventional non-orthogonal multiple access assisted backscatter communication (NOMABC) system without RIS, and demonstrate the feasibility and effectiveness of the proposed algorithm.
	\end{abstract}
 \vspace{-0.5cm}
	\section{Introduction}
Backscatter communication has been actively studied as a low-power, low-complexity and short-range communication technology for Internet of Things (IoT)~\cite{9222571}. The key idea of backscatter communication is to ask an energy-constrained backscatter device (BD) to carry out passive reflection and modulation of a sinusoidal continuous wave sent by a carrier transmitter (CT). Meanwhile, non-orthogonal multiple access (NOMA) has received considerable attention for its great potential to support massive IoT devices and enhance spectrum efficiency~\cite{liu2018non}. NOMA allows multiple users to access the same orthogonal resource block. To support more users and further improve the system performance,  it is natural to consider the combination of backscatter communication and NOMA.

For NOMA assisted backscatter communication (NOMABC) systems, the resource allocation problem was first studied in~\cite{8851217}. The aim was to maximize the minimum throughput among all BDs by jointly optimizing the backscatter time and power reflection coefficients. The secure beamforming problem for the multiple-input single-output (MISO) NOMABC system was considered in~\cite{Yiqing2019}, where the objective was to maximize the outage secrecy rate. 
A new cognitive NOMABC network was proposed in~\cite{9175026}, and the transmit power of the primary user and the reflection coefficients of BDs were jointly optimized. 
To better exploit NOMA in backscatter communication systems, a new reflection coefficient selection criteria was proposed in~\cite{8439079}. To illustrate the proposed criteria, the performance
of the NOMABC system in terms of the average number of bits was analyzed.
 
On the other hand, owing to the capability of smartly reconfiguring the wireless propagation environment, reconfigurable intelligent surfaces (RISs) have received significant attention for their potential to enhance the capacity and coverage of wireless networks~\cite{liu2020reconfigurable}. An RIS is made of electromagnetic (EM) material, which consists of a large number of reconfigurable passive elements. Each elements can reflect the incident signal by appropriately tuning its amplitude and phase. Therefore, RISs have the capability of enhancing the received signal power, overcoming the path loss, and suppress the co-channel interference of the users.

There have been extensive works on RISs and their contributions focus on diverse application
scenarios under different assumptions, such as backscatter communication~\cite{Jia2020,9017956}, NOMA systems~\cite{9167258,guo2020intelligent,fu2019reconfigurable}, and simultaneous wireless information and power transfer (SWIPT) networks~\cite{9133435}. Particularly, in~\cite{9167258}, the subchannel assignment, power allocation, phase shifts and decoding order were optimized jointly by maximizing the achievable sum rate. The downlink RIS assisted NOMA (RIS-NOMA) system over fading channels was considered in~\cite{guo2020intelligent}, where the joint optimization problem over resource allocation and phase shifts was solved by maximizing the average sum rate. For MISO RIS-NOMA systems, the active beamforming and passive beamforming were optimized jointly by minimizing the total transmit power in~\cite{fu2019reconfigurable}. The RIS assisted bistatic backscatter networks was first studied in~\cite{Jia2020}, where the transmit beamforming vector was jointly optimized with the RIS phase shifts. In~\cite{9017956}, the performance of backscatter technology with RIS in terms of the symbol error probability was evaluated.

It is interesting to investigate the promising applications of the RIS technique in NOMABC systems for further performance improvement. To the best of our knowledge, the RIS enhanced NOMABC (RIS-NOMABC) system design has not been studied yet. In this paper, we consider the RIS-NOMABC system. Our objective is to jointly design the power reflection coefficients at the BDs and the phase shifts at the RIS such that the system sum rate is maximized, subject to the minimum quality of service (QoS) requirement for each BDs. To solve the formulated problem, we propose a joint optimization algorithm based on the alternative optimization, successive convex approximation (SCA) and manifold optimization approaches.

Notations: $\mathbb{C}^{M \times 1}$ denotes a complex vector of size \emph{M}; diag(\textbf{x}) denotes a diagonal matrix whose diagonal elements are the corresponding elements in vector \textbf{x}; The $m$-th element of vector $\textbf{x}$ is denoted as $\left[ \mathbf{x} \right] _m$; $\mathbf{x}^{\dagger}$ and ${\mathbf{x}}^{H}$ denote the conjugate and conjugate transpose of vector \textbf{x}, respectively; The notation $\angle x$ denotes the phase of a complex number \emph{x}; $\odot $ denotes the Hadamard products; The function $\mathcal{R}\left( x \right) $ denotes the real part of a complex number \emph{x}; $\mathcal{C}\mathcal{N}\left( 0,\sigma ^2 \right) $ represents a random vector following the distribution of zero mean and $\sigma ^2$ variance;
 \vspace{-0.8cm}
 \section{System Model}
  \begin{figure}
 	\centering
 	\includegraphics[width=2.5in]{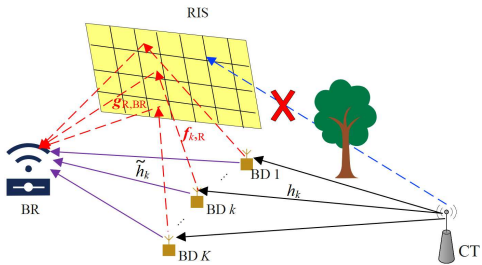}
 	\caption{RIS-NOMABC system model}\label{system_model}
 \end{figure}
We consider a RIS-NOMABC system, as shown in Fig.~\ref{system_model}, which consists of one CT, one RIS, one backscatter receiver (BR), and $K$ BDs. Each of the CT, BR, and BDs is equipped with a single antenna. The RIS is equipped with $Q_{\rm RIS}$ passive reflecting elements. 
The CT transmits sinusoidal carrier signals, and the BDs modulate their information over incident carriers by intelligently changing their load impedances. The BR receives the signals with the aid of the RIS. We assume that the direct CT-RIS link is blocked.

Let $P_{\rm T}$ be the CT's transmit power, and $s_{k}$ be the information symbol of each BDs. Denoting by $w_k$ the power reflection coefficient of the $k$-th BD. The backscattered signal at the $k$-th BD is~\cite{8851217}
\begin{equation}\label{x_k signal}
 \setlength{\abovedisplayskip}{5pt}
\setlength{\belowdisplayskip}{5pt}  
	x_k=h_k \sqrt{w_kP_{\rm T}}s_k
\end{equation} 
 where $h_k\in \mathbb{C}^{1\times 1}$ is the channel from the CT to the $k$-th BD. 
 
 The signal received at the BR is
\begin{equation}\label{signal at BR}  
 \setlength{\abovedisplayskip}{5pt}
 \setlength{\belowdisplayskip}{5pt}
	y=\sum_{k=1}^K{\left( \widetilde{h}_k+\boldsymbol{g}_{\mathrm{R},\mathrm{BR}}^{H}\mathbf{\Theta }\mathbf{f}_{k,\mathrm{R}} \right) h_k\sqrt{w_kP_{\rm T}}s_k} + z,
\end{equation} 
 where $\widetilde{h}_k\in \mathbb{C}^{1\times 1}$, $\mathbf{g}_{\mathrm{R,BR}}\in \mathbb{C}^{Q_{\rm RIS}\times 1}$, $\mathbf{f}_{k,\mathrm{R}}\in \mathbb{C}^{Q_{\rm RIS}\times 1}$ are the channel from the $k$-th BD to the BR, the RIS to the BR, the $k$-th BD to the RIS, respectively, $\mathbf{\Theta }=\mathrm{diag}\left\{ e^{j\theta _1},e^{j\theta _2},\cdots ,e^{j\theta _{Q_{\rm RIS}}} \right\}  $ is the phase-shift matrix of the RIS, $z\sim \mathcal{CN}\left( 0,\sigma^{2} \right)$ is the additive white Gaussian noise (AWGN) with zero mean and variance $\sigma^2$ at the BR.
 
 In conventional uplink NOMA systems, the users with higher channel gains are often decoded earlier at the base station. However, this ordering method cannot be applied in RIS-NOMABC systems, because the combined channel gains can be modified by tuning the RIS phase shifts. The optimal decoding order in the RIS-NOMABC system will be any one of the $K!$ different decoding orders.
Let $\mathcal{D}_k$ denote the decoding order for the signal of BD $k$, where $\mathcal{D}_k=m$ means that the signal of BD $k$ is the $m$-th one to be decoded at the BR. For any two BDs, i.e., BD $j$ and BD $k$, we assume that BD $j$ is decoded after BD $k$, which means that $\mathcal{D}_j>\mathcal{D}_k$.
In addition, the combined channel gains of the two BDs need to satisfy the following condition:
\begin{equation} \label{channel gain condition} 
 \setlength{\abovedisplayskip}{5pt}
\setlength{\belowdisplayskip}{5pt}
\left| \left( \widetilde{h}_k+\boldsymbol{g}_{\mathrm{R},\mathrm{BR}}^{H}\mathbf{\Theta f}_{k,\mathrm{R}} \right) h_k \right|>\left| \left( \widetilde{h}_j+\boldsymbol{g}_{\mathrm{R},\mathrm{BR}}^{H}\mathbf{\Theta f}_{j,\mathrm{R}} \right) h_j \right|
\end{equation}  

According to the NOMA protocol~\cite{liu2018non}, the achievable data rate of BD $k$ is given by
\begin{equation} \label{Rate k} 
 \setlength{\abovedisplayskip}{5pt}
\setlength{\belowdisplayskip}{5pt} 
R_k=\log _2\left( 1+\frac{w_kP_{\rm T}H_k}{\sum_{\mathcal{D}_j>\mathcal{D}_k}{w_jP_{\rm T}H_j}+\sigma ^2} \right) 
\end{equation}  
where $H_k=\left| \left( \widetilde{h}_k+\boldsymbol{g}_{\mathrm{R},\mathrm{BR}}^{H}\mathbf{\Theta f}_{k,\mathrm{R}} \right) h_k \right|^2$ is the the combined channel gain of BD $k$.
 
We aim to maximize the system sum rate of all BDs through appropriate power reflection coefficients at the BDs and phase shfits at the IRS. Therefore, the formulated joint power reflection coefficients and phase shifts optimization problem is given by
\begin{subequations}\label{OP}
 \setlength{\abovedisplayskip}{5pt}
\setlength{\belowdisplayskip}{5pt}
	\begin{align}
	&\underset{\theta _q,w_k}{\max}\sum_{k=1}^K{\log _2\left( 1+\frac{w_kP_{\rm T}H_k}{\sum_{\mathcal{D}_j>\mathcal{D}_k}{w_jP_{\rm T}H_j}+\sigma ^2} \right)}
	, \label{OP:a}
	\\
	&s.t.~\log _2\left( 1+\frac{w_kP_{\rm T}H_k}{\sum_{\mathcal{D}_j>\mathcal{D}_k}{w_jP_{\rm T}H_j}+\sigma ^2} \right) \geqslant R_{k}^{\min}, \label{OP:b} \\
	&   \ \ \ \ \ H_k>H_j, ~{\rm if}~ \mathcal{D}_j>\mathcal{D}_k, k\ne j, \label{OP:c}\\	
	&   \ \ \ \ \ 0\leqslant w_k\leqslant 1, \label{OP:d}\\
	&   \ \ \ \ \ \theta _q\in \left[ 0,2\pi \right] , \label{OP:e} \\
	&   \ \ \ \ \ \mathcal{D}_k\in \mathbb{D}, \label{OP:f}
	\end{align}
\end{subequations}
where $ \mathbb{D}$ is the set of all possible decoding orders, $q=1,2,\cdots ,Q_{\rm RIS}$ and    $k,j=1,2,\cdots ,K$.
The above problem~\eqref{OP} is challenging, not only due to the non-convex objective function
 and constraints, but also due to that the parameters to be optimized are entangled with each other. In the following sections, we will develop an alternative optimization based algorithm to decouple the optimization variables.
  \vspace{-0.4cm}
 \section{Proposed Solution}
  Since the total number of decoding order combinations is a finite value, the optimal system sum rate can be obtained by solving problem~\eqref{OP} with any one of decoding orders at first and selecting the maximum objective function’s value among all decoding orders. Without loss of generality, we set $\mathcal{D}_k=k$. Then, the system sum rate of the RIS-NOMABC system can be re-expressed as
 \begin{equation} \label{sum rate} 
  \setlength{\abovedisplayskip}{5pt}
 \setlength{\belowdisplayskip}{5pt}
 \begin{split}
 R_{\mathrm{sum}} &=\sum_{k=1}^K{\log _2\left( 1+\frac{w_kP_{\rm T}H_k}{\sum_{j=k+1}^K{w_jP_{\rm T}H_j}+\sigma ^2} \right)}
 \\&
\overset{\left( a \right)}{=}\log _2\left( 1+\frac{\sum_{k=1}^K{w_kP_{\rm T}H_k}}{\sigma ^2} \right), 
 \end{split}
 \end{equation} 
 where $\left( a \right)$ come from the fact that the terms inside the brackets of the system sum rate expression forms a telescoping product.
 
 \begin{remark}
 	The system sum rate in~\eqref{sum rate} is independent of the decoding order. However, different decoding orders
 	result in different achievable rates of each BDs and different feasible regions of the power reflection coefficients.
 \end{remark}
 
Thus, the system sum rate maximization problem in~\eqref{OP} is reduced to 
\begin{subequations}\label{OP_1}
	\begin{align}
	&\underset{\theta _q,w_k}{\max}\log _2\left( 1+\frac{\sum_{k=1}^K{w_kP_{\rm T}H_k}}{\sigma ^2} \right) , \label{OP_blocked:a}
	\\
	&s.t.~\log _2\left( 1+\frac{w_kP_{\rm T}H_k}{\sum_{j=k+1}^K{w_jP_{\rm T}H_j}+\sigma ^2} \right) \geqslant R_{k}^{\min}, \label{OP_blocked:b} \\
	& \ \ \ \ \ H_k>H_j,~{\rm if}~j>k, \label{OP_blocked:c}\\	
	&   \ \ \ \ \ \eqref{OP:d},~\eqref{OP:e}. \label{OP_blocked:d}
	\end{align}
\end{subequations} 

Problem~\eqref{OP_1} is a non-convex optimization problem. To make it tractable, we first decouple it into two sub-problems, i.e., power reflection coefficients optimization and phase shifts optimization. Then, we solve them alternatively.
 \vspace{-0.5cm}
 \subsection{Power Reflection Coefficients Optimization}
First, we focus our attention on the power reflection coefficients optimization problem. Since the log function is a monotonic increasing function, maximizing $\log _2\left( 1+\frac{\sum_{k=1}^K{w_kP_{\rm T}H_k}}{\sigma ^2} \right)$ is equivalent to maximizing $\sum_{k=1}^K{w_kH_k}$. Then, the power reflection coefficients optimization problem with fixed phase shifts $\left\{ \theta _q \right\} $ in~\eqref{OP_1} can be expressed as follows
 \begin{subequations}\label{PRC_blocked}
 \setlength{\abovedisplayskip}{5pt}
\setlength{\belowdisplayskip}{5pt}
 	\begin{align}
 	&\underset{w_k}{\max}~\sum_{k=1}^K{w_kH_k}
 	, \label{PRC_blocked:a}
 	\\
 	&s.t.~w_k\geqslant \frac{r_{k}^{\min}\left( P_{\rm T}\sum_{j=k+1}^K{w_jH_j}+\sigma ^2 \right)}{P_tH_k}
 	, \label{PRC_blocked:b} \\
 	&   \ \ \ \ \ \eqref{OP:d}. \label{PRC_blocked:c}
 	\end{align}
 \end{subequations} 

Clearly, the above problem is convex, and can be solved using standard convex algorithms. However, the standard approach does not exploit the specific structure of problem~\eqref{PRC_blocked}. In the following, we derive the closed-form optimal solution of the power reflection coefficients for problem~\eqref{PRC_blocked}.

Since the last BD $K$ suffers no inference generated from the other BDs, therefore the minimum QoS constraint in~\eqref{PRC_blocked:b} for BD $K$ is given by:
 \begin{equation} \label{min p_K}
  \setlength{\abovedisplayskip}{5pt}
 \setlength{\belowdisplayskip}{5pt} 
w_K\geqslant \frac{r_{K}^{\min}\sigma ^2}{P_{\rm T}H_K}=w_{K}^{\rm LB}
 \end{equation}

According to the definition of the achievable data rate in~\eqref{Rate k}, the lower bound power reflection coefficient $w_K^{\rm LB}$ should be utilized to reduce the interference from BD $K$ to the other BDs. Therefore, for user $K-1$, we have
 \begin{equation} \label{min p_K-1} 
  \setlength{\abovedisplayskip}{5pt}
 \setlength{\belowdisplayskip}{5pt}
	\begin{split}
		w_{K-1}& \geqslant \frac{r_{K-1}^{\min}\left( w_KP_{\rm T}H_K+\sigma ^2 \right)}{P_{\rm T}H_{K-1}}
		\\&
		\geqslant \frac{\sigma ^2r_{K-1}^{\min}\left( r_{K}^{\min}+1 \right)}{P_{\rm T}H_{K-1}}=w_{K-1}^{\rm LB}
	\end{split}
\end{equation} 

Likewise, it is easy to extend the above inequality to all BDs. According to~\cite{8892657}, we have 
 \begin{equation} \label{min p_k} 
w_k\geqslant \frac{\sigma ^2r_{k}^{\min}}{P_{\rm T}H_k}\prod_{j=k+1}^K{\left( r_{j}^{\min}+1 \right)}=w_{k}^{\rm LB}
\end{equation} 
where $\prod_{j=K+1}^K{\left( r_{j}^{\min}+1 \right)}=1
$.
 
\begin{theorem}
For BD $1$, the optimal power reflection coefficient $w_{1}^{*}=1$. For BD $k$ $\left( k\geqslant 2 \right) $, if the optimal power reflection coefficient for BD $m$ $\left( m<k \right) $ are all equal to 1, i.e., $w_m=1 $, then the optimal solution of power reflection coefficient for BD $K$ is	
 \begin{equation} \label{optimal power k} 
 w_{k}^{*}=\min \left\{ 1,w_{k}^{\mathrm{UB}} \right\} 
\end{equation} 
with the upper bound $w_{k}^{\mathrm{UB}}$ of $ w_{k}$ is defined as
 \begin{equation} \label{UB power k} 
w_{k}^{\mathrm{UB}}=\underset{m}{\min}\left\{ \frac{1}{H_k}\left( \frac{H_m}{r_{k}^{\min}}-\widetilde{H}_k-\frac{\sigma ^2}{P_{\mathrm{T}}} \right) \right\} ,
\end{equation} 
where $\widetilde{H}_k=\sum_{n=m+1}^{k-1}{H_n}+\sum_{j=k+1}^K{w_{j}^{\mathrm{LB}}H_j}
$.

If $w_{k}^{*}=w_{k}^{\mathrm{UB}}$, the optimal power reflection coefficient of the other BDs are $w_{j}^{*}=w_{j}^{\mathrm{LB}}$ for $j>k$.

Proof: Similar proof can be found in~\cite{8892657}.
\end{theorem}
  \vspace{-0.5cm}
 \subsection{Phase Shifts Optimization}
Define $\mathbf{b}_k=\left[ h_k\boldsymbol{g}_{\mathrm{R},\mathrm{BR}}^{H}\mathrm{diag}\left( \mathbf{f}_{k,\mathrm{R}} \right) \,\, h_k\widetilde{h}_k \right] ^H $ and let $\mathbf{v}=\left[ e^{j\theta _1}\cdots e^{j\theta _{Q_{\rm RIS}}}\,\,1 \right] ^T $ be the passive beamforming vector, then the phase shifts optimization problem with fixed  $\left\{ w _k \right\} $ in~\eqref{OP_1} can be written as
 \begin{subequations}\label{PSO}
 \setlength{\abovedisplayskip}{5pt}
\setlength{\belowdisplayskip}{5pt}
 	\begin{align}
 	&\underset{\mathbf{v}}{\max}\sum_{k=1}^K{w_k\left| \mathbf{b}_{k}^{H}\mathbf{v} \right|^2}, \label{PSO:a}
 	\\
 	&s.t.~w_k\left| \mathbf{b}_{k}^{H}\mathbf{v} \right|^2\geqslant r_{k}^{\min}\left( \sum_{j=k+1}^K{w_j\left| \mathbf{b}_{j}^{H}\mathbf{v} \right|^2}+\frac{\sigma ^2}{P_{\mathrm{T}}} \right) , \label{PSO:b} \\
 	&   \ \ \ \ \left| \mathbf{b}_{k}^{H}\mathbf{v} \right|^2>\left| \mathbf{b}_{j}^{H}\mathbf{v} \right|^2, ~{\rm if}~ j>k, \label{PSO:c} \\
 	&   \ \ \ \ \left| \left[ \mathbf{v} \right] _q \right|=1,q=1,2,\cdots ,Q_{\mathrm{RIS}}+1.  \label{PSO:d}  
 	\end{align}
 \end{subequations}
 
 The non-convex constraints~\eqref{PSO:b}, ~\eqref{PSO:c}, and~\eqref{PSO:d} make the solving problem~\eqref{PSO} difficult. In the following subsections, we propose a suboptimal algorithm to solve problem~\eqref{PSO}. First, we introduce the auxiliary variables $\left\{ a_k \right\} $, which satisfy
 \begin{equation} \label{auxiliary variables}  
  \setlength{\abovedisplayskip}{5pt}
   \setlength{\belowdisplayskip}{5pt}
     a_k=\mathbf{b}_{k}^{H}\mathbf{v}
\end{equation}
 
 Substituting~\eqref{auxiliary variables} into problem~\eqref{PSO}, then we have
\begin{subequations}\label{PSO_1}
	 \setlength{\abovedisplayskip}{5pt}
	 \setlength{\belowdisplayskip}{5pt}
	\begin{align}
	&\underset{\mathbf{v},a_k}{\max}\sum_{k=1}^K{w_k\left| a_k \right|^2}	, \label{PSO_1:a}	\\
	&s.t.~w_k\left| a_k \right|^2\geqslant r_{k}^{\min}\left( \sum_{j=k+1}^K{w_j\left| a_j \right|^2}+\frac{\sigma ^2}{P_{\mathrm{T}}} \right) 
	, \label{PSO_1:b} \\
	&   \ \ \ \ \left| a_k \right|^2>\left| a_j \right|^2, ~{\rm if}~ j>k, \label{PSO_1:c} \\
	&   \ \ \ \ a_k=\mathbf{b}_{k}^{H}\mathbf{v},\label{PSO_1:d} \\
	&   \ \ \ \ \left| \left[ \mathbf{v} \right] _q \right|=1.\label{PSO_1:e} 
 	\end{align}
\end{subequations}

By utilizing the penalty-based method, we first convert the equality constraints in~\eqref{PSO_1:d} into quadratic functions and then add them as a penalty term in the objective function of~\eqref{PSO_1}. Thus, we have  
\begin{subequations}\label{PSO_2}
  \setlength{\abovedisplayskip}{5pt}
  \setlength{\belowdisplayskip}{5pt}
	\begin{align}
	&\underset{\mathbf{v},a_k}{\min}-\sum_{k=1}^K{w_k\left| a_k \right|^2}+\mu \left( \sum_{k=1}^K{\left| a_k-\mathbf{b}_{k}^{H}\mathbf{v} \right|^2} \right) 
	, \label{PSO_2:a}
	\\
	&s.t.~\eqref{PSO_1:b},~\eqref{PSO_1:c},~\eqref{PSO_1:e} .  \label{PS_blocked_2:e}  
	\end{align}
\end{subequations}
where $\mu$ denotes the penalty coefficient used for penalizing the violation of equality constraints~\eqref{PSO_1:d}.

Problem~\eqref{PSO_2} is still a non-convex optimization problem due to the non-convex objective function as well as non-convex constraints. In the following, we solve the auxiliary variables $\left\{ a_k \right\} $  and passive beamforming vector $\mathbf{v}$ respectively.
 \subsubsection{Auxiliary variables $\left\{ a_k \right\} $ optimization}
 With given $\left\{ w_k \right\}$ and $\mathbf{v}$, the auxiliary variables optimization problem is formulated as
 \begin{subequations}\label{AVO}
 	\setlength{\abovedisplayskip}{5pt}
 	\setlength{\belowdisplayskip}{5pt}
 	\begin{align}
 	&\underset{a_k}{\min}-\sum_{k=1}^K{w_k\left| a_k \right|^2}+\mu \left( \sum_{k=1}^K{\left| a_k-\mathbf{b}_{k}^{H}\mathbf{v} \right|^2} \right) 
 	, \label{AVO-a}
 	\\
 	&s.t.~\eqref{PSO_1:b},~\eqref{PSO_1:c}
 	\end{align}
 \end{subequations}

Problem~\eqref{AVO} is a non-convex problem, due to the non-convex constraints~\eqref{PSO_1:b} and~\eqref{PSO_1:c}. To deal with the non-convexity, the SCA method can be used. At the point $ a_{k}^{\left( t_1 \right)}$, the first order approximation of $\left| a_k \right|^2$ is 
 \begin{equation} \label{SCA} 
\left| a_k \right|^2\geqslant 2\mathcal{R}\left( \left( a_{k}^{\left( t_1 \right)} \right) ^{\dagger}a_k \right) -\left| a_{k}^{\left( t_1 \right)} \right|^2
\\
=f_{\rm SCA}\left( a_k,a_{k}^{\left( t_1 \right)} \right)
 \end{equation} 
 
 By substituting the above approximation into problem~\eqref{AVO}, then we have 
 \begin{subequations}\label{AVOSCA}
 \setlength{\abovedisplayskip}{5pt}
\setlength{\belowdisplayskip}{5pt}
	\begin{align}
	&\underset{a_k}{\min}-\sum_{k=1}^K{w_kf_{\rm SCA}\left( a_k,a_{k}^{\left( t_1 \right)} \right)}+\mu \left( \sum_{k=1}^K{\left| a_k-\mathbf{b}_{k}^{H}\mathbf{v} \right|^2} \right) 
	, \label{AVOSCA:a}
	\\
	&s.t.~w_kf_{\mathrm{SCA}}\left( a_k,a_{k}^{\left( t_1 \right)} \right) \geqslant r_{k}^{\min}\left( \sum_{j=k+1}^K{w_j\left| a_j \right|^2}+\frac{\sigma ^2}{P_{\mathrm{T}}} \right) , \label{AVO:b}  	\\
	&   \ \ \ \ f_{\rm SCA}\left( a_k,a_{k}^{\left( t_1 \right)} \right) >\left| a_j \right|^2,~{\rm if}~ j>k. \label{AVO:c}  	
	\end{align}
\end{subequations}
 
It is noted that problem~\eqref{AVOSCA} is a convex problem, which can be efficiently solved via standard convex problem solvers such as CVX~\cite{cvx}. \textbf{Algorithm~\ref{SCA algorithm}} summarizes the proposed SCA-based algoirthm to solve problem~\eqref{AVO}. According to~\cite{Marks}, the proposed SCA-based algorithm converges to a stationary point that satisfies the Karush-Kuhn-Tucker (KKT) conditions.
  \vspace{-0.3cm}
\begin{algorithm}
	\caption{The proposed SCA-based algorithm to solve problem~\eqref{AVO}}
	\label{SCA algorithm}
	\begin{algorithmic}[1]
		\STATE  Initialize feasible $a_{k}^{\left( 0 \right)} $ and set the iteration index $t_1 = 0$.
		\REPEAT
		\STATE Update $a_{k}^{\left( t_1+1 \right)}$ by solving problem~\eqref{AVOSCA} with $a_{k}^{\left( t_1 \right)}
		$;	
		\STATE Update $t_1 = t_1 + 1$;	  
		\UNTIL the objective function of problem~\eqref{AVOSCA} converges
		\STATE   \textbf{Output}: optimal $\left\{ a_k \right\} 
		$.
	\end{algorithmic}
\end{algorithm}  
 
 In \textbf{Algorithm~\ref{SCA algorithm}}, the initial feasible points $a_{k}^{\left( 0 \right)} $ are needed. Usually, it is difficult to find the feasible points. In the following, we formulate a feasibility problem and propose a novel feasible initial points searching algorithm. By introducing an infeasibility indicator $x\ge 0$, the feasibility problem in the ${t_2}$-th iteration is given as
 \begin{subequations}\label{feasible problem}
 	 \setlength{\abovedisplayskip}{5pt}
 	\setlength{\belowdisplayskip}{5pt}
  	\begin{align}
 	& \min_{a_k,x} x,    \\
 	&{\rm s.t.} \ w_kf_{\mathrm{SCA}}\left( a_k,a_{k}^{\left( t_2 \right)} \right) +x\geqslant r_{k}^{\min}\left( \sum_{j=k+1}^K{w_j\left| a_j \right|^2}+\frac{\sigma ^2}{P_{\mathrm{T}}} \right) , \\
 	& \ \ \ \ f_{\rm SCA}\left( a_k,a_{k}^{\left(  t_2 \right)} \right) + x>\left| a_j \right|^2, \\
 	& \ \ \ \ \ x\geqslant 0,
 	\end{align}
 \end{subequations}
 where $x$ denotes how far the corresponding constrains in problem~\eqref{AVOSCA} are from being satisfied.
 
Problem~\eqref{feasible problem} is also a convex optimization problem and the proposed feasible initial points searching algorithm to solve problem~\eqref{feasible problem} is similarly as \textbf{Algorithm~\ref{SCA algorithm}}. Due to the space limit, we omit the details of the proposed feasible initial points searching algorithm
 \subsubsection{Passive beamforming vector $\mathbf{v}$ optimizaiton}
  With given $\left\{ w_k \right\}$ and  $\left\{ a_k \right\}$, the passive beamforming vector optimization problem can be written as
\begin{subequations}\label{PBVO}
 \setlength{\abovedisplayskip}{5pt}
\setlength{\belowdisplayskip}{5pt}
	\begin{align}
	&\underset{\mathbf{v}}{\min} \ \sum_{k=1}^K{\left| a_k-\mathbf{b}_{k}^{H}\mathbf{v} \right|^2}, \label{PBVO:a} \\
	&s.t.~ \left| \left[ \mathbf{v} \right] _q \right|=1,q=1,2,\cdots ,Q_{\mathrm{RIS}}. 	\label{PBVO:b} 
	\end{align}
\end{subequations}

The main difficulty to solve problem~\eqref{PBVO} is the non-convex unit modulus constraint~\eqref{PBVO:b}. To the best of our knowledge, there is no general approach to solve such optimization problem optimally. In the following, the manifold optimization approach~\cite{Yashuai} is utilized to solve problem~\eqref{PBVO}. We first define the manifold space for the constraint~\eqref{PBVO:b} in problem~\eqref{PBVO} as
 \begin{equation} \label{manifold space} 
  \setlength{\abovedisplayskip}{5pt}
 \setlength{\belowdisplayskip}{5pt}
\mathbb{V}=\left\{ \mathbf{v}\in \mathbb{C}^{Q_{\mathrm{RIS}}\times 1}|\left| \left[ \mathbf{v} \right] _1 \right|=\cdots =\left| \left[ \mathbf{v} \right] _{Q_{\mathrm{RIS}}} \right|=1 \right\} 
 \end{equation} 
 
According to the notion of manifold optimization, problem~\eqref{PBVO} can be reformulated as:
 \begin{equation} \label{manifold optimization} 
  \setlength{\abovedisplayskip}{5pt}
 \setlength{\belowdisplayskip}{5pt}
\underset{\mathbf{v}\in \mathbb{V}}{\min}~f\left( \mathbf{v} \right) =\sum_{k=1}^K{\left| a_k-\mathbf{b}_{k}^{H}\mathbf{v} \right|^2}
\end{equation}   
  
The main idea of the manifold optimization approach is to apply the gradient descent algorithm
in the manifold space. In particular, the main steps of the manifold optimization approach is composed of the following steps at the $t_3$-th iteration:  
  \begin{enumerate}
		\item Calculate the Euclidean gradient: the Euclidean gradient of $f\left( \mathbf{v}^{\left( t_3 \right)} \right)$ at $\mathbf{v}^{\left( t_3 \right)}$ can be computed by
		 \begin{equation} \label{Euclidean gradient} 
		  \setlength{\abovedisplayskip}{5pt}
		 \setlength{\belowdisplayskip}{5pt}
		\nabla _{\mathbf{v}}f\left( \mathbf{v}^{\left( t_3 \right)} \right) =2\left( \sum_{k=1}^K{\mathbf{b}_k\mathbf{b}_{k}^{H}} \right) \mathbf{v}^{\left( t_3 \right)}-2\sum_{k=1}^K{a_k\mathbf{b}_k}
		 \end{equation}
		 \item  Calculate the Riemannian gradient: The Riemannian gradient is one
		tangent vector (direction) with the decrease of the objective function over the manifold space. For manifold space $\mathbb{V}$, the tangent space at $\mathbf{v}^{\left( t_3 \right)}$ is given by:
		\begin{equation} \label{Tangent space} 
			\mathbb{T}\left( \mathbf{v}^{\left( t_3+1 \right)} \right) =\left\{ \mathbf{z}\in  \mathbb{C}^{Q_{\mathrm{RIS}}\times 1}|\mathcal{R}\left( \mathbf{z}\odot \left( \mathbf{v}^{\left( t_3 \right)} \right) ^{\dagger} \right) =\textbf{0 }\right\} 
		\end{equation}
		Then, the Riemannian gradient of $f\left( \mathbf{v}^{\left( t_3 \right)} \right)$ at $\mathbf{v}^{\left( t_3 \right)}$ can be obtained by orthogonally projecting the Eculidean gradient $\nabla _{\mathbf{v}}f\left( \mathbf{v}^{\left( t_3 \right)} \right)$ on to the tangent space $\mathbb{T}\left( \mathbf{v}^{\left( t_3+1 \right)} \right) $ given by
		\begin{equation} \label{Riemannian gradient} 
		 \setlength{\abovedisplayskip}{5pt}
		\setlength{\belowdisplayskip}{5pt}
		\begin{array}{l}
		\nabla _{\mathbb{V}}f\left( \mathbf{v}^{\left( t_3 \right)} \right)\\
		 = \nabla _{\mathbf{v}}f\left( \mathbf{v}^{\left( t_3 \right)} \right) -\mathcal{R}\left( \left( \nabla _{\mathbf{v}}f\left( \mathbf{v}^{\left( t_3 \right)} \right) \right) ^{\dagger}\odot \mathbf{v}^{\left( t_3 \right)} \right) \odot \mathbf{v}^{\left( t_3 \right)}
		\end{array}
		\end{equation}
	 \item Update the current point $\widetilde{\mathbf{v}}^{\left( t_3 \right)}
	 $: After we obtain the Riemannian gradient, the current	$\widetilde{\mathbf{v}}^{\left( t_3 \right)}
	 $ in the tangent space $\mathbb{T}\left( \mathbf{v}^{\left( t_3+1 \right)} \right) $ is updated as
	 \begin{equation} \label{Update the current point} 
	  \setlength{\abovedisplayskip}{5pt}
	 \setlength{\belowdisplayskip}{5pt}
	\widetilde{\mathbf{v}}^{\left( t_3 \right)}=\mathbf{v}^{\left( t_3 \right)}-\lambda \nabla _{\mathbb{V}}f\left( \mathbf{v}^{\left( t_3 \right)} \right) 
	 \end{equation} 
	 where $\lambda >0$ is a constant step size, which is selected to satisfy $\lambda \leqslant \frac{1}{\lambda _{\max}\left( \sum_{k=1}^K{\mathbf{b}_k\mathbf{b}_{k}^{H}} \right)}
	 $ with $\lambda _{\max}\left( \sum_{k=1}^K{\mathbf{b}_k\mathbf{b}_{k}^{H}} \right) 
	 $ be the largest eigenvalue of the matrix $\sum_{k=1}^K{\mathbf{b}_k\mathbf{b}_{k}^{H}}
	 $. It should be noticed that the update point $\widetilde{\mathbf{v}}^{\left( t_3 \right)}
	 $ is still in the tangent space $\mathbb{T}\left( \mathbf{v}^{\left( t_3+1 \right)} \right) $ and it leaves the manifold space.
	\item Retraction mapping: To map the updated point $\widetilde{\mathbf{v}}^{\left( t_3 \right)}
	$ onto the manifold space $\mathbb{V}$, a Retraction mapping operator is needed. Finally, the point $\mathbf{v}^{\left( t_3+1 \right)}$ updated by using the Retraction mapping operator is given by
	 \begin{equation} \label{Retraction}
	 \begin{split}
	 \mathbf{v}^{\left( t_3+1 \right)} & =\mathcal{R_M}\left( -\lambda \nabla _{\mathbb{V}}f\left( \mathbf{v}^{\left( t_3 \right)} \right) \right) \\
	 &  
	 =\frac{\mathbf{v}^{\left( t_3 \right)}-\lambda \nabla _{\mathbb{V}}f\left( \mathbf{v}^{\left( t_3 \right)} \right)}{\left\| \mathbf{v}^{\left( t_3 \right)}-\lambda \nabla _{\mathbb{V}}f\left( \mathbf{v}^{\left( t_3 \right)} \right) \right\|}=\widetilde{\mathbf{v}}^{\left( t_3 \right)}\odot \frac{1}{\widetilde{\mathbf{v}}^{\left( t_3 \right)}}
	 \end{split}
	 \end{equation}
 \end{enumerate}

\begin{figure}
    	 \setlength{\abovecaptionskip}{-0.4cm}
     	\setlength{\belowcaptionskip}{-0.5cm}  
	\centering
	\includegraphics[width=2.5in]{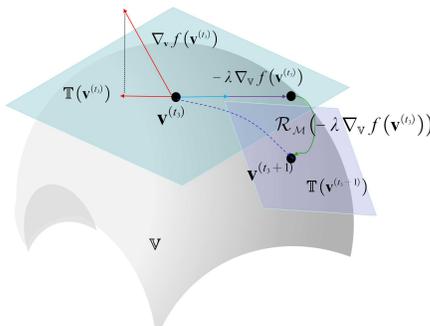}
	\caption{The geometric interpretation of the manifold optimization algorithm}\label{Fig2}
\end{figure}

 The proposed manifold optimization algorithm to solve problem~\eqref{PBVO} is summarized in \textbf{Algorithm~\ref{MO algorithm}} and is also illustrated geometrically in Fig.~\ref{Fig2}. According to~\cite{Yashuai,8706630},\textbf{ Algorithm~\ref{MO algorithm}} is guaranteed to converge to the point where the gradient of the objective function is zero.
  \vspace{-0.3cm}
\begin{algorithm}
	\caption{The proposed manifold optimization algorithm to solve problem~\eqref{PBVO}}
	\label{MO algorithm}
	\begin{algorithmic}[1]
		\STATE  Initialize $\mathbf{v }^{\left( 0 \right)}$ and set the iteration index ${t_3} = 0$.
		\REPEAT
		\STATE  ${t_3} = {t_3} + 1$;
		\STATE  Calculate the Euclidean gradient $\nabla _\textbf{v}f\left( \mathbf{v}^{\left( t_3 \right)} \right)  	$ at $\mathbf{v}^{\left( t_3 \right)}$ using~\eqref{Euclidean gradient}; 
		\STATE  Calculate the Riemannian gradient $\nabla _{\mathbb{V}} f\left( \mathbf{v}^{\left( t_3 \right)} \right) 	$ using~\eqref{Riemannian gradient}; 
		\STATE Update the current point $\widetilde{\mathbf{v}}^{\left( t_3 \right)}
		$ using~\eqref{Update the current point}; 
		\STATE Update $\mathbf{v}^{\left( t_3 +1 \right)}$ ) using the Retraction mapping operator according to~\eqref{Retraction};
		\UNTIL  $\left| f\left( \mathbf{v}^{\left( t_3 +1 \right)} \right) -f\left( \mathbf{v}^{\left( t_3 \right)} \right) \right|
		$ converges
		\STATE   \textbf{Output}: optimal $\mathbf{v}$.
	\end{algorithmic}
\end{algorithm} 
 \vspace{-0.5cm}
\subsection{Proposed Algorithm, Convergence and Complexity} 
 To facilitate the understanding of the proposed algorithm to solve problem~\eqref{OP_1}, we summarize it in \textbf{Algorithm~\ref{proposed algorithm}}. The objective value of problem~\eqref{OP_1} is monotonically non-decreasing after each iteration and the system sum rate is upper bounded. Therefore, the proposed algorithm is guaranteed to converge. The complexity of \textbf{Algorithm~\ref{proposed algorithm}} mainly depends on \textbf{Algorithm~\ref{SCA algorithm}} and \textbf{Algorithm~\ref{MO algorithm}} with complexities $O\left( t_{1}^{\max}K^3 \right) $ and $O\left( t_{3}^{\max}\left( Q_{\mathrm{RIS}}+1 \right) ^2 \right) 
 $, respectively, where $t_{1}^{\max}$ and $t_{3}^{\max}$ are the iteration numbers of \textbf{Algorithm~\ref{SCA algorithm}} and \textbf{Algorithm~\ref{MO algorithm}} required for convergence.
   \vspace{-0.3cm}
\begin{algorithm}
	\caption{The proposed algorithm to solve problem~\eqref{OP_1}}
	\label{proposed algorithm}
	\begin{algorithmic}[1]
		\STATE  Initialize $\left\{ \theta _{q}^{\left( 0 \right)} \right\} $ and set the iteration index ${t_4} = 0$.
		\REPEAT
		\STATE  update $ w_{k}^{\left( t_4+1 \right)}$ according to \textbf{Theorem 1} with ${\theta }_q^{\left( {t_4 } \right)}$;
		\STATE  update $a_{k}^{\left( t_4+1 \right)}$ by \textbf{Algorithm~\ref{SCA algorithm}} with  $ w_{k}^{\left( t_4+1 \right)}$ and ${\theta }_q^{\left( {t_4 } \right)}$; 
		\STATE  update $\mathbf{v}^{\left( {t_4+1 } \right)}$ by \textbf{Algorithm~\ref{MO algorithm}} with $ w_{k}^{\left( t_4+1 \right)}$ and  $a_{k}^{\left( t_4+1 \right)}$;
		\STATE  caculate ${\theta }_q^{\left( {t_4+1 } \right)}=\angle \left[ \mathbf{v}^{\left( t_4 +1\right)} \right] _q$;
		\STATE  ${t_4} = {t_4} + 1$;
		\UNTIL {the objective value of problem~\eqref{OP_1} converges.}
		\STATE   \textbf{Output}: optimal $\left\{ w_k \right\} $ and $\left\{ \theta_q \right\} $.
	\end{algorithmic}
\end{algorithm} 
 \vspace{-0.5cm}
\section{Numerical Results}
Here, the performance of the proposed algorithm is evaluated through numerical simulations. Assume that the CT, RIS and BR are located at coordinates (0 m, 10 m), (65 m, 10 m) and (70 m, 10 m), respectively. The BDs are randomly and uniformly placed in the area between coordinates (40m, 0m) and (50m, 0m). The distance-dependent path loss is modeled as $P\left( d \right) =\rho \left( d \right) ^{-\alpha}$, where \emph{d} is the link distance, $\alpha$ is the path loss exponent, $\rho=-30~\rm dB$ is the path loss at the reference distance of 1 m. The path loss exponents of the CT-BD, BD-BR, BD-RIS, and RIS-BR links are set as 2.5, 2.5, 2.1 and 2.1, respectively. To model the small-scale fading for all channnels involved, we adopt Rician fading, which is given by $f_{\rm Rician}=\sqrt{\frac{\kappa}{1+\kappa}}f_{\rm Rician}^{\text{LoS}}+\sqrt{\frac{1}{1+\kappa}}f_{\rm Rician}^{\text{NLoS}}$, where $\kappa$ is the Rician factor, ${{{f}}_{\rm Rician}^{\rm Los}}$ and ${{{f}}_{\rm Rician}^{\rm NLoS}}$ are the line-of-signt (LoS) component and non-LoS (NLoS) component, respectively. We set the Rician factor $\kappa =3$ for BD-RIS and RIS-BR links and $\kappa =0$ for other communication links. We assume that the number of BDs is $K=3$ and the noise power is ${\sigma ^2} =-114~\rm dBm$.

\begin{figure}
  	\setlength{\belowcaptionskip}{-0.1cm}  
	\centering
	\includegraphics[width=2.5in]{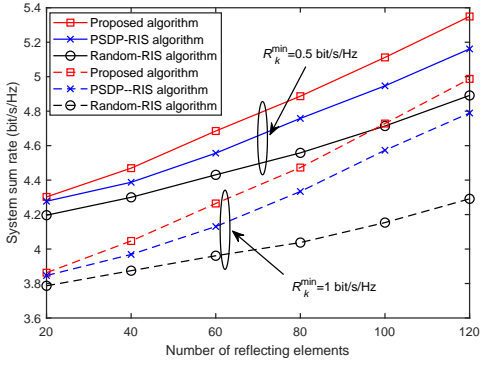}
	\caption{System sum rate versus the number of reflecting elements $Q_{\rm RIS}$, $P_{\rm T}=35 ~ \rm dBm$}\label{SumRate_Q}
\end{figure}

In order to validate the effectiveness of our proposed algorithm, two benchmark schemes are considered, namely, PSDP-RIS algorithm and Random-RIS algorithm. For the PSDP-RIS algorithm, the phase shifts optimization problem~\eqref{PSO} is solved by the penalty function based semidefinate programming (PSDP) algorithm~\cite{lin2019joint}. For the Random-RIS algorithm, the phase shifts are selected randomly. The optimal power reflection coefficients for both algorithms are obtained according to \textbf{Thereom 1}. Fig.~\ref{SumRate_Q} depicts the impact of the number of RIS reflecting elements on the system sum rate. As expected, we can see from Fig.~\ref{SumRate_Q} that the system sum rate achieved by the three algorithms increases as the number of reflecting elements increase because a larger number of RIS reflecting elements leads to a higher passive array gains. In addition, we observe that our proposed algorithm has the best performance.
 \vspace{-0.1cm}

\begin{figure}
 	\setlength{\belowcaptionskip}{-0.1cm}  
	\centering
	\includegraphics[width=2.5in]{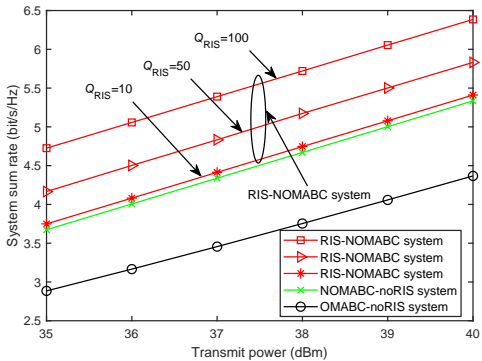}
	\caption{System sum rate versus the transmit power $P_{\rm T}$, $R_k^{\rm min}=1 ~\rm bit/s/Hz$}\label{SumRate_PT}
\end{figure}

Fig.~\ref{SumRate_PT} shows the system sum rate versus the transmit power $P_{\rm T}$. We observe that the achieved system sum rate of all schemes increases with $P_{\rm T}$. In particular, the NOMA-based systems outperform the OMA assisted backscatter communication without RIS (OMABC-noRIS) system, since all users can be served simultaneously in the NOMA-based systems. Furthermore, our proposed RIS-NOMABC system with large number of RIS reflecting elements significantly outperforms the NOMABC without RIS (NOMABC-noRIS) system, which reveals that the application of RIS to the NOMABC system can further improve the system sum rate.
 \vspace{-0.1cm}

\begin{figure}
  \setlength{\belowcaptionskip}{-0.1cm}  
	\centering
	\includegraphics[width=2.5in]{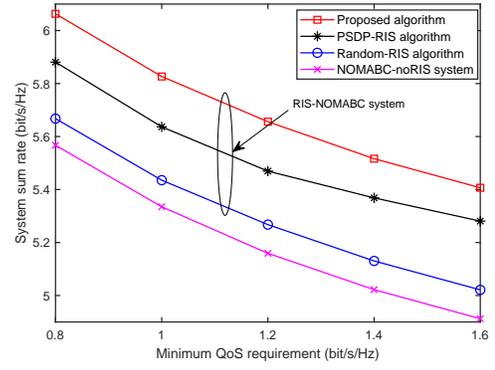}
	\caption{System sum rate versus the minimum QoS requirement $R_k^{\rm min}$,  $Q_{\rm RIS}=50$, $P_{\rm T}=40~\rm dBm$}\label{SumRate_Rmin}
\end{figure}

In Fig~\ref{SumRate_Rmin}, we present the system sum rate versus the minimum QoS requirement $R_k^{\rm min}$. It is observed that our proposed algorithm has the best performance than the other schemes, which can also be observed in Fig.~\ref{SumRate_Q} and Fig.~\ref{SumRate_PT}. In addition, the schemes for the proposed RIS-NOMABC system significantly outperforms the scheme for NOMABC-noRIS system.
 \vspace{-0.1cm}
 \vspace{-0.3cm}
\section{Conclusion}
 \vspace{-0.1cm}
This paper proposed an RIS enhanced NOMA backscatter assisted communication system. The joint power reflection coefficients and phase shifts optimization was investigated. The non-convex problem was solved by the alternative optimization, successive convex approximation and manifold optimization. Our numerical results showed that the proposed algorithm has a better performance than benchmark algorithms. Furthermore, our results revealed that the proposed RIS-NOMABC system with a large number of RIS reflecting elements can achieve significantly system sum rate gain compared with conventional NOMABC and OMABC systems. This insight provides useful guidelines for the practical RIS implementation.



 \vspace{-0.3cm}
 \bibliographystyle{IEEEtran}
 \bibliography{zjkbib}

\end{document}